\documentclass[12pt]{article}

\usepackage[dvipdfmx]{graphicx,color}
\usepackage{amsmath,amssymb}
\usepackage{bm}
\usepackage{wrapfig}
\usepackage{cancel}
\begin{document}
\title{
\begin{flushright}
\ \\*[-80pt] 
\begin{minipage}{0.2\linewidth}
\normalsize
HUPD1603
IPMU16-0046
 \\*[25pt]
\end{minipage}
\end{flushright}
{\Large \bf 
Occam's Razor in Lepton Mass Matrices \\
{\large \bf- The Sign of Universe's Baryon Asymmetry - }  
\\*[14pt]}}

\author{ 
\centerline{
Yuya~Kaneta$^{1,}$\footnote{E-mail address: kaneta@muse.sc.niigata-u.ac.jp}, ~
~Yusuke~Shimizu$^{2,3,}$\footnote{E-mail address: yu-shimizu@hiroshima-u.ac.jp},} \\
\\
\centerline{Morimitsu~Tanimoto$^{4}$ ~ and ~
~Tsutomu~T.~Yanagida$^{5}$
}
\\*[12pt]
\centerline{
\begin{minipage}{\linewidth}
\begin{center}
$^1${\it \normalsize
Graduate~School~of~Science~and~Technology,~Niigata University, \\
Niigata~950-2181,~Japan} \\
$^{2}${\it \normalsize
Graduate School of Science, Hiroshima University, \\
 Higashi-Hiroshima, 739-8526, Japan} \\
$^{3}$ {\it \normalsize
School~of~Physics,~KIAS,~Seoul~130-722,~Republic~of~Korea} \\
$^4${\it \normalsize
Department~of~Physics,~Niigata~University,~Niigata~950-2181,~Japan} \\
$^5${\it \normalsize 
Kavli~IPMU,~TODIAS,~University~of~Tokyo,~Kashiwa~277-8583,~Japan}
\end{center}
\end{minipage}}
\\*[65pt]}

\date{
\centerline{\small \bf Abstract}
\begin{minipage}{0.9\linewidth}
\medskip 
\medskip 
\small
We discuss the neutrino mass matrix based on the Occam's Razor approach
in the framework of the seesaw mechanism.
We impose four zeros in the Dirac neutrino mass matrix, which give
the minimum number of parameters needed for the observed neutrino
masses and lepton mixing angles,
while the charged lepton  mass matrix and the right-handed Majorana neutrino mass
matrix are taken to be real diagonal ones.
The low-energy neutrino mass matrix has only seven physical parameters.
We show successful predictions for the mixing angle $\theta_{13}$  and
the CP violating phase $\delta_{CP}$ with the normal mass hierarchy of neutrinos
 by using the experimental data on the neutrino mass squared differences,
the mixing angles $\theta_{12}$ and $\theta_{23}$. 
The most favored region of  $\sin\theta_{13}$ is around
$0.13\sim 0.15$, which is completely consistent with the observed value.
The CP violating phase  $\delta_{CP}$ is favored to be close to $\pm
\pi/2$. We also discuss the Majorana phases as well as the effective
neutrino mass for the neutrinoless double-beta decay $m_{ee}$, which is
around $7\sim 8$ meV.
It is extremely remarkable that we can perform a ``complete experiment"
to determine the low-energy neutrino mass matrix, since we have only
seven physical parameters in the neutrino mass matrix. 
In particular, two CP violating phases in the neutrino mass matrix are directly given by two CP
violating phases at high energy. Thus, assuming the leptogenesis we can
determine the sign of the cosmic baryon in the universe from the low-energy
experiments for the neutrino mass matrix.
\end{minipage}
}

\begin{titlepage}
\maketitle
\thispagestyle{empty}
\end{titlepage}


\section{Introduction}

The standard model has been  well established by the  discovery of the Higgs boson.
However, the origin and structure of quark and lepton flavors are still unknown in spite of the remarkable success of the standard model.
Therefore, underlying physics for the masses and mixing
of quarks and leptons is one of the fundamental problems in particle physics.
Actually, a number of models have been proposed based on flavor symmetries, but there is no convincing model at present.

On the other hand, the neutrino oscillation experiments are going on a new step to reveal the CP violation in the lepton sector.
The T2K experiment has confirmed the neutrino oscillation in the $\nu_\mu \to\nu_e$ appearance events \cite{Abe:2013hdq},
which may provide us a new information of the CP violation in the lepton sector. Recent NO$\nu$A experimental data \cite{Adamson:2016tbq} also indicate the CP violation in  the neutrino oscillation. Thus, various informations are now available
to discuss Yukawa matrices in the lepton sector.

Recently, the Occam's Razor approach was proposed
 to investigate the neutrino mass matrix \cite{Harigaya:2012bw}
in the case of two heavy right-handed neutrinos. Because of tight constraints it was shown that only the inverted mass hierarchy for the
neutrinos is consistent with the present experimental data. The quark sector was also successfully discussed in this approach \cite{Tanimoto:2016rqy} and we found a nice prediction of the Cabibbo angle, for instance.

In this paper, we discuss the seesaw mechanism \cite{seesaw} with the  three right-handed heavy Majorana neutrinos, predicting the normal  mass hierarchy of the light neutrinos.
We impose four  zeros in the Dirac neutrino  mass matrix, which give
the minimum number of parameters needed  for the observed neutrino
masses and lepton mixing angles in the normal mass hierarchy of neutrinos
 \cite{Branco:2007nb,Choubey:2008tb}.
Here, the charged lepton  mass matrix and the right-handed Majorana neutrino mass matrix
are taken to be real diagonal ones.
The Dirac neutrino mass matrix is given with five complex parameters.
Among them,  three phases are removed by the phase redefinition of the three left-handed neutrino fields. The remained two phases are removed by the  field-phase rotation of the right-handed neutrinos. Instead, these two phases appear in the right-handed Majorana neutrino mass matrix. After integrating the heavy right-handed neutrinos, we obtain a mass matrix of the light neutrino,
which contains five real parameters and two CP violating phases.

In the present Occam's Razor approach with the four zeros of the Dirac neutrino mass matrix,
we show the successful predictions of the mixing angle $\theta_{13}$  and
the CP violating phase $\delta_{CP}$ with the normal mass hierarchy of neutrinos. We also discuss the Majorana phases and the effective neutrino mass of the neutrinoless double-beta decay.

It is extremely remarkable that we can perform a ``complete experiment"
to determine the low-energy neutrino mass matrix \cite{Branco:2002ie}, since we have only
seven physical parameters in the neutrino mass matrix. In particular, two CP violating
phases in the neutrino mass matrix are directly related to two CP
violating phases at high energy. Thus, assuming the leptogenesis, we can
determine the sign of cosmic baryon in the universe only from the low-energy experiments for the neutrino mass matrix \cite{Frampton:2002qc}.

 In section 2, we show a viable  Dirac  neutrino  mass matrix with four zeros,  where
  we take the real diagonal basis  of the charged lepton mass matrix and the right-handed
  Majorana neutrino mass matrix. We also present qualitative discussions of our parameters
  in order to reproduce the two large mixing angles of neutrino flavors.
  In section 3, we show numerical results for our  mass matrix.
  The  summary is devoted in section 4.
  In Appendix, we show  parameter relations in our mass matrix.

\section{Neutrino mass matrix}
\label{sec:lepton-model}

On the standpoint of Occam's Razor approach
\cite{Harigaya:2012bw,Tanimoto:2016rqy},
 we discuss the  neutrino mass matrix in the framework of the seesaw
mechanism without assuming any symmetry.
We take the real diagonal basis of the charged lepton mass matrix and
the right-handed Majorana neutrino mass matrix as:
\begin{equation}
M_E=
\begin{pmatrix}
m_e & 0 & 0 \\
0 & m_\mu& 0 \\
0 & 0 & m_\tau
\end{pmatrix}_{LR} \ ,  \quad \quad
M_R=
\begin{pmatrix}
M_1 & 0 & 0 \\
0 & M_2 & 0 \\
0 & 0 & M_3
\end{pmatrix}_{RR}.
\end{equation}
We reduce the number of free parameters in the Dirac neutrino mass
matrix by putting zero at several elements in the matrix.
The four zeros of the Dirac neutrino mass matrix give us the minimum
number of parameters to reproduce the observed neutrino masses and
lepton mixing angles. This is what we call the Occam's Razor approach.

The successful  Dirac neutrino mass matrix with four zeros
\footnote{Other four zero textures may be available for the lepton mixing.
Those will be discussed comprehensively in the future work.}
 is given as
\begin{equation}
m_D=
\begin{pmatrix}
0 & A & 0 \\
A' & 0 & B \\
0 & B' & C
\end{pmatrix}_{LR},
\end{equation}
which  has  five complex parameters.
\footnote{$A'=0$ corresponds to the case discussed in ref. \cite{Harigaya:2012bw}.
 Thus, five zero textures are not excluded.}
The three phases can be removed by the phase rotation of the three left-handed neutrino fields.  This phase redefinition does not affect the lepton mixing matrix because the charged
lepton mass matrix is diagonal and the phases are absorbed in the three right-handed charged lepton fields. In order to get the real matrix for the Dirac neutrino mass matrix,
 the remained two phases are removed by the phase rotation of the two right-handed neutrino
fields.
Instead,  the right-handed  Majorana neutrino mass matrix becomes complex diagonal one as follows:
\begin{equation}
M_R=
\begin{pmatrix}
M_1 e^{-i\phi_A} & 0 & 0 \\
0 & M_2 e^{-i\phi_B} & 0 \\
0 & 0 & M_3
\end{pmatrix}_{RR} = M_0
\begin{pmatrix}
\frac{1}{k_1} e^{-i\phi_A} & 0 & 0 \\
0 & \frac{1}{k_2} e^{-i\phi_B} & 0 \\
0 & 0 & 1
\end{pmatrix}_{RR},
\end{equation}
where $M_0\equiv M_3$, $k_1=M_3/M_1$ and $k_2=M_3/M_2$.
 We obtain the left-handed Majorana neutrino mass matrix after integrating out the heavy right-handed neutrinos,
 \begin{align}
m_\nu =m_D M_R^{-1}m_D^T 
=\frac{1}{M_0}
\begin{pmatrix}
A^2 k_2 e^{i\phi_B} & 0 &  A B' k_2 e^{i\phi_B}\\
0 & A'^2 k_1 e^{i\phi_A}+ B^2 & B C\\
A B' k_2 e^{i\phi_B} & B C  &  B'^2 k_2 e^{i\phi_B} +C^2
\end{pmatrix} ,
\label{matrix}
\end{align}
 in which there are ten parameters apparently. 
However,  it is expressed in terms of  seven parameters by the rescaling of parameters.
Let us replace parameters  by introducing  new parameters 
$a$, $b$, $c$, $k'_1$ and $k'_2$ as,
\begin{equation}
A=\sqrt{M_0 k'_2}\  a, \quad A'=\sqrt{M_0 k'_1}\  a, \quad B=\sqrt{M_0}\ b, 
\quad B'=\sqrt{M_0 k'_2}\  b, \quad C=\sqrt{M_0}\ c.
\end{equation}
Then, the neutrino mass matrix is written as 
 \begin{align}
m_\nu 
=
\begin{pmatrix}
a^2 K_2 e^{i\phi_B} & 0 &  ab K_2 e^{i\phi_B}\\
0 & a^2 K_1 e^{i\phi_A}+ b^2 & bc\\
ab K_2 e^{i\phi_B} & bc  &  b^2 K_2 e^{i\phi_B} +c^2
\end{pmatrix} ,
\label{numatrix}
\end{align}
where
\begin{equation}
K_1=k'_1 k_1=\left (\frac{A'B'}{AB} \right )^2\frac{M_3}{M_1}\ ,
\qquad \qquad K_2=k'_2 k_2
=\left (\frac{B'}{B} \right )^2\frac{M_3}{M_2}\ .
\end{equation}
Finally, the neutrino mass matrix is expressed by five real parameters,
$a, b,c, K_1,K_2$ and two phases $\phi_A, \phi_B$.
Since we can input five experimental data of neutrinos,  the  mass squared differences
$\Delta m^2_{\rm atm}$, $\Delta m^2_{\rm sol}$ and three lepton mixing angles $\theta_{23}$,
$\theta_{12}$ and $\theta_{13}$, there remains two free parameters.
Those two parameters are determined by the Dirac CP violating phase $\delta_{CP}$ and the effective neutrino mass $m_{ee}$ for the neutrinoless double-beta decay \cite{Branco:2002ie}.

Here we comment on the concern with the texture zero analysis of the left-handed neutrino mass matrix \cite{Frampton:2002yf}. Actually, some two zero textures of the left-handed neutrino mass matrix are consistent with the recent  data \cite{Singh:2016qcf}.
On the other hand, our neutrino mass matrix of Eq.(\ref{numatrix}) is a zero one texture.
The two zero textures are never realized without the tuning among parameters
as seen in  Eq.(\ref{matrix}) since  we start with the seesaw mechanism of the neutrino masses,
 in which we take
 the right-handed Majorana neutrino mass matrix to be diagonal \cite{Kageyama:2002zw}.
Although there are seven parameters in the neutrino mass matrix in Eq.(\ref{numatrix}),
we can give clear predictions at the large $K_1$ and $K_2$, which corresponds to
the large mass  hierarchy among the right-handed Majorana neutrinos.

We can obtain the eigenvectors by solving the eigenvalue equation of Eq.(\ref{numatrix}).
The mass eigenvalues are expressed by  $a, b,c, K_1,K_2$ and $\phi_A, \phi_B$
as seen in Appendix.
And then,  we get the lepton mixing matrix, so called the 
Maki-Nakagawa-Sakata (MNS) matrix $U_{\text{MNS}}$~\cite{Maki:1962mu,Pontecorvo:1967fh}.
 It is expressed in terms of 
three mixing angles $\theta _{ij}$ $(i,j=1,2,3;~i<j)$, the CP violating Dirac phase 
$\delta _{CP}$ and two Majorana phases $\alpha$ and $\beta$  as
\begin{align}
U_\text{MNS} \equiv 
\begin{pmatrix}
c_{12} c_{13} & s_{12} c_{13} & s_{13}e^{-i\delta _{CP}} \\
-s_{12} c_{23} - c_{12} s_{23} s_{13}e^{i\delta _{CP}} & 
c_{12} c_{23} - s_{12} s_{23} s_{13}e^{i\delta _{CP}} & s_{23} c_{13} \\
s_{12} s_{23} - c_{12} c_{23} s_{13}e^{i\delta _{CP}} & 
-c_{12} s_{23} - s_{12} c_{23} s_{13}e^{i\delta _{CP}} & c_{23} c_{13}
\end{pmatrix}
\begin{pmatrix}
e^{i\alpha} & 0& 0 \\ 0 & e^{i\beta} & 0 \\ 0 & 0 & 1
\end{pmatrix}, 
\label{MNS}
\end{align}
where $c_{ij}$ and  $s_{ij}$ denote $\cos \theta _{ij}$ and $\sin \theta _{ij}$, respectively.

There is a CP violating observable, the Jarlskog invariant $J_{CP}$ \cite{Jarlskog:1985ht},
which is derived from the following relation:
\begin{align}
&i {\cal C}\equiv [M_\nu M_\nu^\dagger, M_E M_E^\dagger]  \ , \nonumber \\
& \det {\cal C}= -2 J_{CP} (m_3^2-m_2^2) (m_2^2-m_1^2) (m_1^2-m_3^2) (m_\tau^2-m_\mu^2) (m_\mu^2-m_e^2) (m_e^2-m_\tau^2) \ ,
\label{Jcp}
\end{align}
where $m_1$, $m_2$ and $m_3$ are neutrino masses with real numbers.
The predicted one is expressed in terms of the parameters of the mass matrix elements as:
\begin{align}
J_{CP}\simeq\frac{1}{2} F \frac{1}{(\Delta m^2_{\rm atm})^2 \Delta m^2_{\rm sol}}  \ ,
\end{align}
where 
\begin{eqnarray}
&&F=2 a^2b^4 c^2 K_2^2 \ 
  \{ b^4K_2\sin\phi_B+a^4K_1K_2 \sin(\phi_A-\phi_B) +\nonumber \\
 &&a^2c^2(K_1\sin \phi_A-K_2\sin \phi_B)+
a^2b^2K_2(K_1\sin(\phi_A+\phi_B)-K_2\sin 2\phi_B-\sin\phi_B)
 \} \ .
\end{eqnarray}
We can extract  $\sin \delta _{CP}$ from  $J_{CP}$
by using the following relation among mixing angles, the Dirac phase and $J_{CP}$ :
\begin{equation}
\sin \delta _{CP}=J_{CP}/(s_{23}c_{23}s_{12}c_{12}s_{13}c_{13}^2) \ .
\end{equation}

The Majorana phases $\alpha$ and $\beta$ are obtained
after diagonalizing the neutrino mass matrix of  Eq.(\ref{numatrix}) as follows:
\begin{equation}
 U_{\rm MNS}^\dagger m_\nu  U^*_{\rm MNS} = {\rm diag} 
\left \{ m_1 ,\  m_2 , \ m_3 \right \} \ .
\end{equation}
Then, we can estimate the effective mass which appears in the neutrinoless double-beta decay   as 
\begin{equation}
m_{ee}=c_{13}^2 c_{12}^2 e^{2i\alpha} m_1 + c_{13}^2 s_{12}^2 e^{2i\beta}m_2+ 
s_{13}^2 e^{-2i\delta_{CP}} m_3 \ .
\end{equation}

The neutrino mass matrix of Eq.(\ref{numatrix}) becomes a simple one at the $K_1$  and $K_2$
large limit with $b^2 K_2$ being finite.
This case corresponds to the large hierarchy of the right-handed neutrino mass ratios $M_3/M_1$
and $M_3/M_2$.
Then,  the magnitudes of our parameters are  estimated  qualitatively
  to reproduce the two large mixing angles $\theta_{23}$  and $\theta_{12}$.
 At first, impose the  maximal mixing of $\theta_{23}$. Then,  the $(2,3)$  element
of Eq.(\ref{numatrix})  should be
 comparable to the $(3,3)$ one, so that the cancellation must be realized 
 between two terms in the $(3,3)$ element, and then we have:
\begin{equation}
K_2\sim \frac{c^2}{b^2}  \ , \qquad   \phi_B\sim \pm \pi \ .
\label{kk1}
\end{equation}
The $(2,3)$ element of Eq.(\ref{numatrix}) is also comparable to the $(2,2)$ one, which is dominated by
the first term $a^2 K_1 \exp(i\phi_A)$ at the  large $K_1$.
So, we get 
\begin{equation}
 K_1\sim \frac{bc}{a^2}  \ .
 \label{kk2}
\end{equation}
At the next step, we impose the large $\theta_{12}$, which requires
the  $(1,3)$ element of Eq.(\ref{numatrix}) to be comparable to $(2,2)$ within a few factor, therefore, we get
\begin{equation}
 a K_1\sim b K_2 r \   ,  \quad (r=2\sim 3) \ .
 \label{kk3}
\end{equation}
By combining Eqs.(\ref{kk1}), (\ref{kk2}), (\ref{kk3}),
we obtain
\begin{equation}
 a c r\sim b^2  \ , \qquad  K_1 \sim \left (\frac{c}{b} \right )^3 \ , 
\qquad K_2\sim \left (\frac{c}{b} \right )^2\ , \qquad K_1^2\sim K_2^3\  r^4\ ,\quad (r=2\sim3)\ .
 \label{kk4}
\end{equation}
Actually, those relations are well satisfied in the numerical result at the large $K_1$.
Then, $\theta_{13}$ becomes rather large, roughly, order of $\sin\theta_{12}/r$
since  the $(1,3)$ element of Eq.(\ref{numatrix}) is comparable to  $(2,3)$ 
within a factor of two or three.
Thus, the seizable mixing angle  $\theta_{13}$ is essentially derived
in this textures when the observed mixing angles $\theta_{23}$ and $\theta_{12}$
 are input. This situation is well reproduced in our numerical result.
 
 Furthermore, we expect the large  CP violating phase $\delta_{CP}$ in this discussion.
 As shown in Eq.(\ref{kk1}), the real part of the $(3,3)$ element of Eq.(\ref{numatrix})
 is significantly suppressed in order to reproduce the almost maximal mixing of  
 $\theta_{23}$.
 Then,   the imaginary part of the  $(3,3)$ element is relatively enhanced
  even if  $\phi_B$ is close to $\pm 180^\circ$.
  Actually, $\phi_B\simeq \pm 175^\circ$ leads to the  $\delta_{CP}\simeq \pm 90^\circ $
  in the numerical analysis of the next section.


\section{Numerical analysis}

Let us discuss the numerical result with the normal mass hierarchy of neutrinos.
At the first step, we constrain the real parameters  $a, b,c, K_1,K_2$ 
and two phases  $\phi_A, \phi_B$ by inputting the
 experimental data of  $\Delta m^2_{\rm atm}$
 and  $\Delta m^2_{\rm sol}$ with $90\%$ C.L. into the relations of Eq.(\ref{massrelations}) 
in Appendix.
 By removing $c$, $\phi_A$ and $\phi_B$ for a fixed $m_1$, which is varied 
 in the region of $m_1=0\sim \sqrt{\Delta m^2_{\rm sol}}$ , there remains 
 four parameters $a, b, K_1$ and $K_2$. 

\begin{figure}[b!]
\begin{minipage}[]{0.45\linewidth}
\includegraphics[width=8cm]{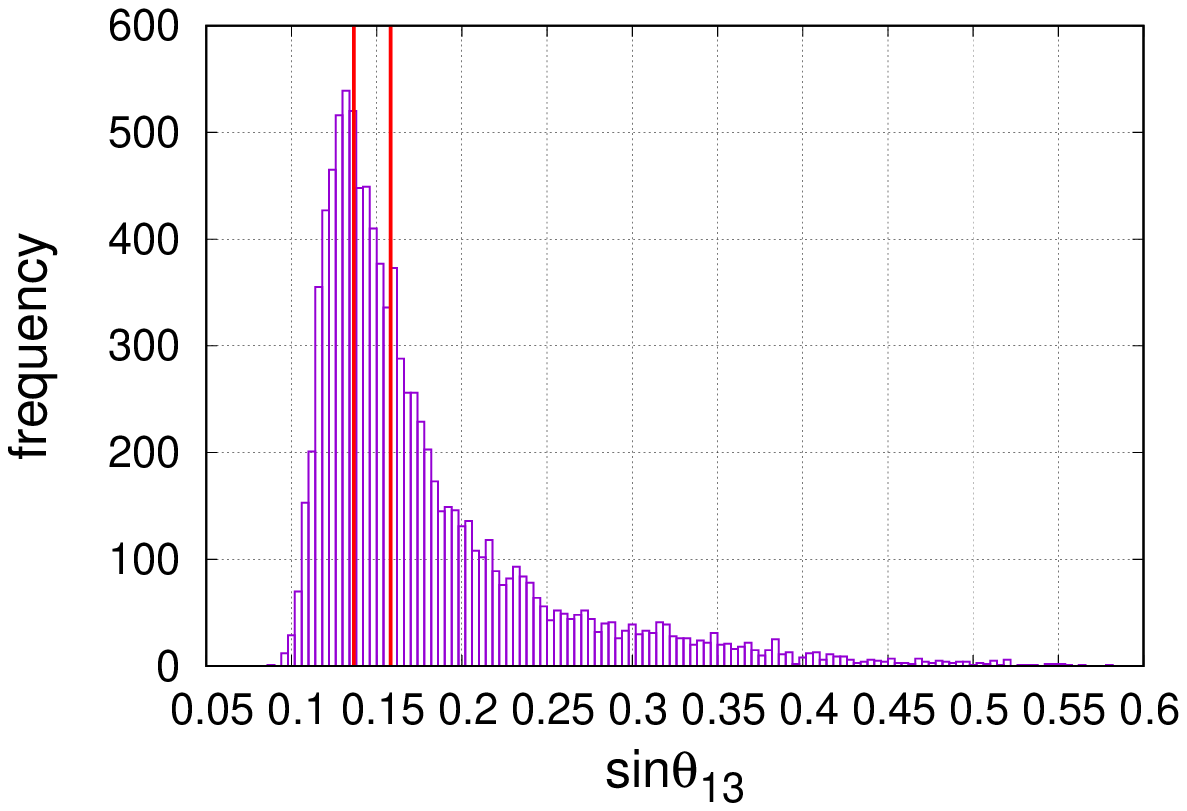}
\caption{The frequency distribution of the predicted  $\sin\theta_{13}$ at $K_1=1-5000$
by inputting the data of  $\theta_{12}$ and $\theta_{23}$.
Here the vertical red lines denote the experimental data  with $3\sigma$. }
\end{minipage}
\hspace{5mm}
\begin{minipage}[]{0.45\linewidth}
\includegraphics[width=8cm]{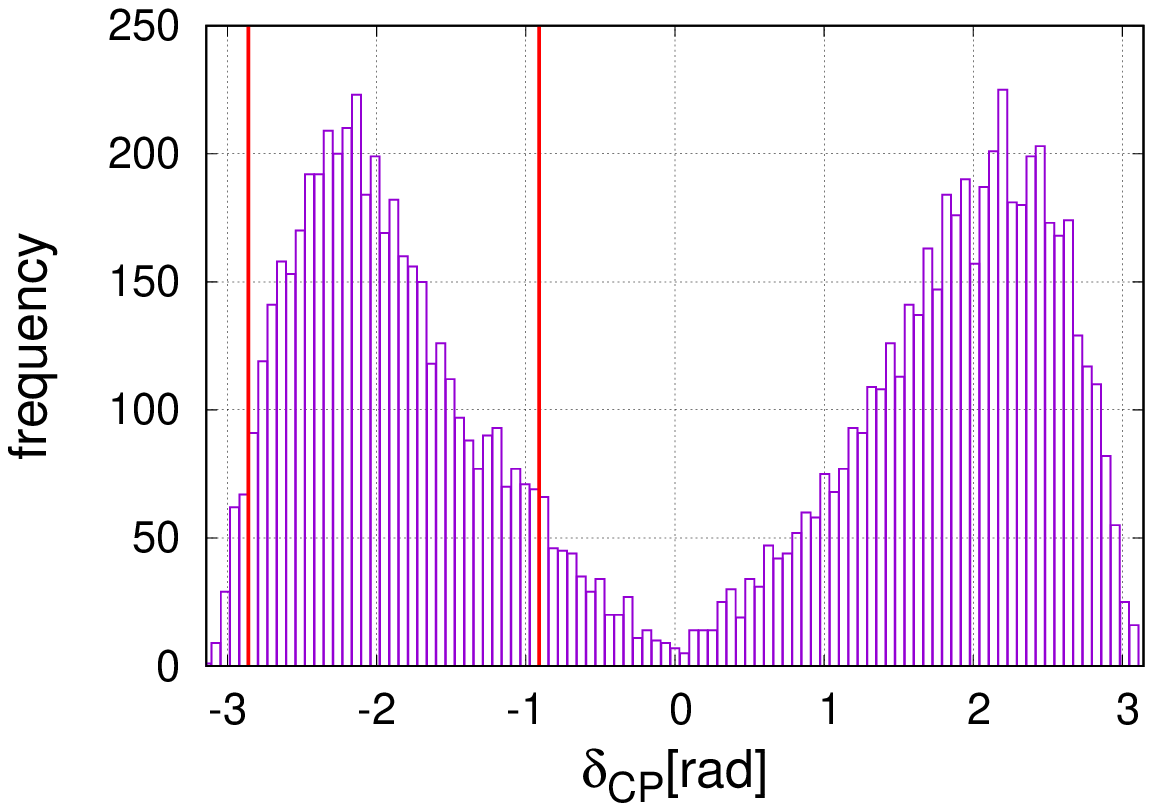}
\caption{ The frequency distribution of the predicted $\delta_{CP}$ at $K_1=1-5000$
by inputting the data of  $\theta_{12}$ and $\theta_{23}$.
Here  the vertical red lines denote the  NO$\nu$A  allowed region
 with $1\sigma$.}
\end{minipage}
\end{figure}
\begin{figure}[h!]
\begin{minipage}[]{0.45\linewidth}
  \includegraphics[width=8cm]{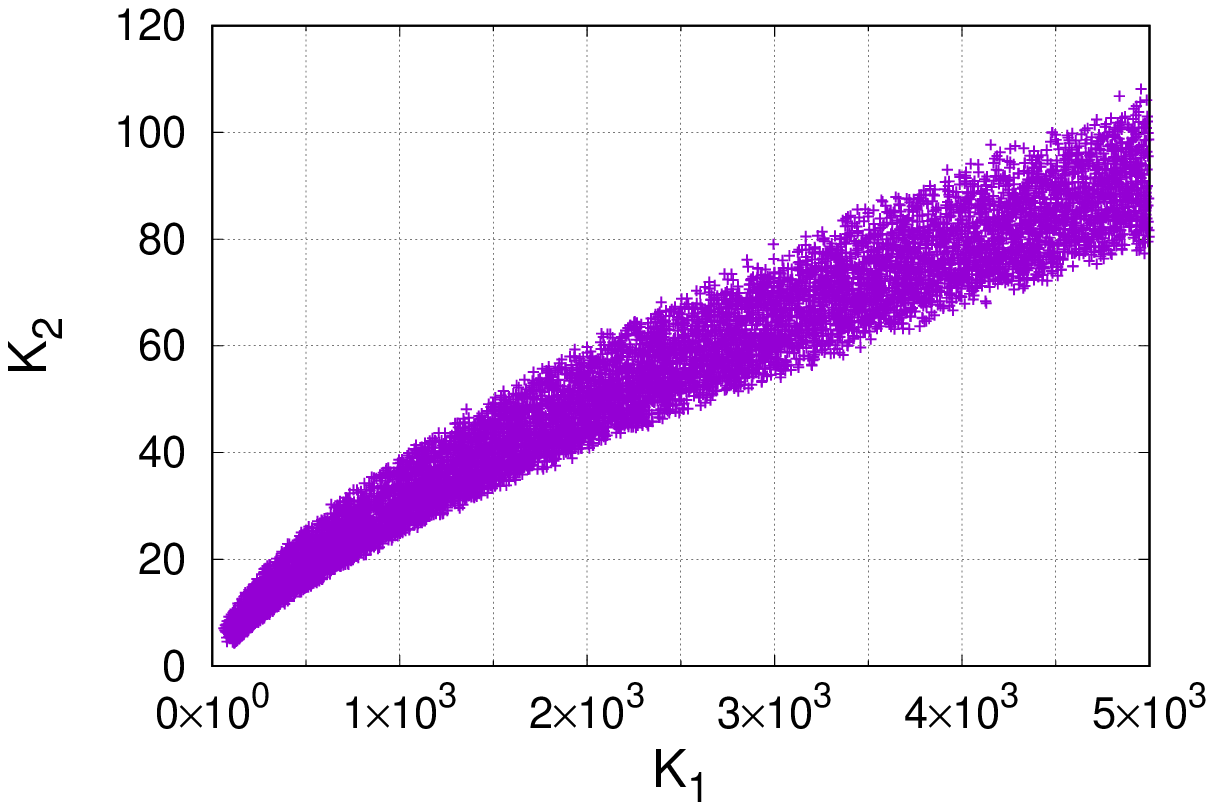}
\caption{The allowed region  on the $K_1-K_2$ plane at  $K_1=1-5000$
by inputting the data of three mixing angles.}
\end{minipage}
\hspace{5mm}
\begin{minipage}[]{0.45\linewidth}
\includegraphics[width=8cm]{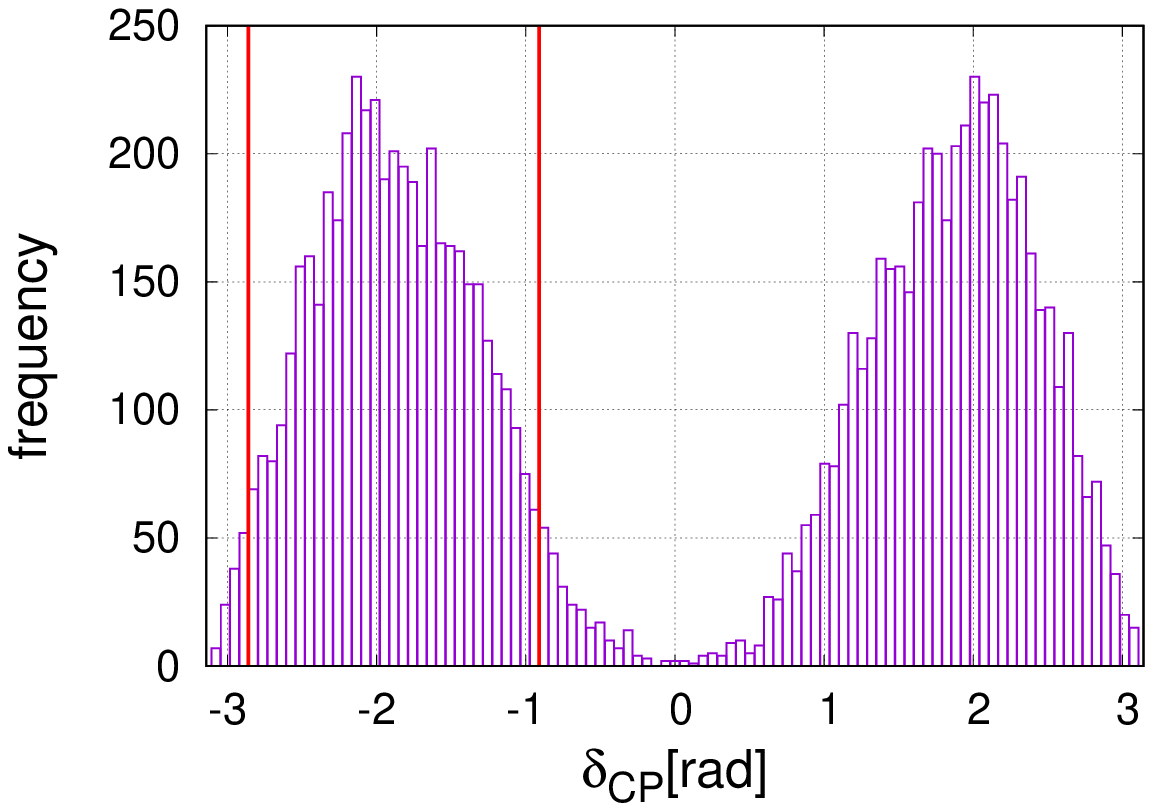}
\caption{The frequency distribution of the predicted $\delta_{CP}$ at $K_1=1-5000$
by inputting the data of three mixing angles.
Here  the vertical red lines denote the  NO$\nu$A  allowed region
 with $1\sigma$.}
\end{minipage}
\end{figure}

At the second step,  we scan them in the following regions 
by generating random numbers in the liner scale  as follows:
\begin{equation}
K_1=[1\sim 10^6], \quad  K_2=[1\sim 10^4], 
\quad  a=[0\sim 0.03]\ {\rm eV^{1/2}}, \quad  b=[0\sim 0.2]\ {\rm eV^{1/2}}.
\end{equation}
They are constrained by the experimental data of the  lepton mixing angles.
And then, we predict
 $\delta_{CP}$, $m_{ee}$, Majorana phases $\alpha$ and $\beta$. 
 The input data are given as follows \cite{Gonzalez-Garcia:2015qrr}:
\begin{eqnarray}
&&\Delta m_{\rm atm}^2= 2.457 \pm 0.047 \times 10^{-3}{\rm eV}^2 \ , \qquad
\Delta m_{\rm sol}^2= 7.50^{+0.19}_{-0.17} \times 10^{-5}{\rm eV}^2 \ , \nonumber \\
&&\nonumber \\
&&\sin^2\theta_{12}=0.304^{+0.013}_{-0.012}\ , \quad 
\sin^2\theta_{23}=0.452^{+0.052}_{-0.028}\ , \quad  \sin^2\theta_{13}=0.0218\pm 0.0010 \ ,
\label{data}
\end{eqnarray}
where we adopt these data with the error-bar of $90\%$ C.L in our calculations. We assume the normal mass hierarchy of neutrinos.
Actually, we have not found the inverted mass hierarchy, in which the three lepton mixing angles  are consistent with the observed values in our numerical calculations. Thus, we consider that the normal mass hierarchy is a prediction in the present model as long as there is no extreme fine tuning of the parameters.

Let us  show the result for $K_1=1-5000$.
By inputting the data of the two mixing angles $\theta_{12}$ and $\theta_{23}$, 
we present the frequency distribution of the predicted  $\sin\theta_{13}$ in Fig.1,
where the vertical red lines denote the experimental data of Eq.(\ref{data}) with $3\sigma$ range.
The peak is  within the  experimental data for $3\sigma$ range.
It is remarked that $\sin\theta_{13}\simeq 0.14$ is most favored.
 This prediction is understandable as discussed  below Eq.(\ref{kk4}).
We also present the frequency distribution of the predicted value of $\delta_{CP}$
in Fig.2, where the vertical red lines denote the  NO$\nu$A experimental allowed region
 with $1\sigma$ range,
which is obtained by the method of Library Event Matching (LEM) \cite{Adamson:2016tbq}.
We see that $\delta_{CP}$ is favored to be around $\pm 2$ radian, which is consistent with
the T2K \cite{Abe:2013hdq} and  NO$\nu$A data for $1\sigma$ range.

 If we add the constraint of the experimental data of $\theta_{13}$, 
 the predictions become rather clear.
By input of the experimental data of $\theta_{13}$,   we obtain the allowed region  on the $K_1-K_2$ plane in Fig.3.
As $K_1$ increases, the $K_2$ also increases gradually.  
This behavior is expected  in  Eq.(\ref{kk4}).
We present the frequency distribution of the predicted value of $\delta_{CP}$
in Fig.4. 
The peak of the distribution is still around $\pm 2$ radian, but the distribution becomes 
rather sharp compared with the one in Fig.2.

Let us discuss the $K_1$ dependence of $\delta_{CP}$,
which is shown in Fig.5.
In the region of $K_1={\cal O}(1-100)$,  the predicted $\delta_{CP}$ is distributed broader.
As $K_1$ increases, the predicted region becomes narrow gradually.
And then, it becomes consistent with the  NO$\nu$A experimental allowed region
 with $1\sigma$ range at the high $K_1$.

We also predict the effective neutrino mass $m_{ee}$, which appears
 in the amplitude of the neutrinoless double-beta decay.
In Fig. 6, we present the frequency distribution of $m_{ee}$.
The favored  $m_{ee}$ is around $7$ meV.

\begin{figure}[t!]
\begin{minipage}[]{0.45\linewidth}
\includegraphics[width=8cm]{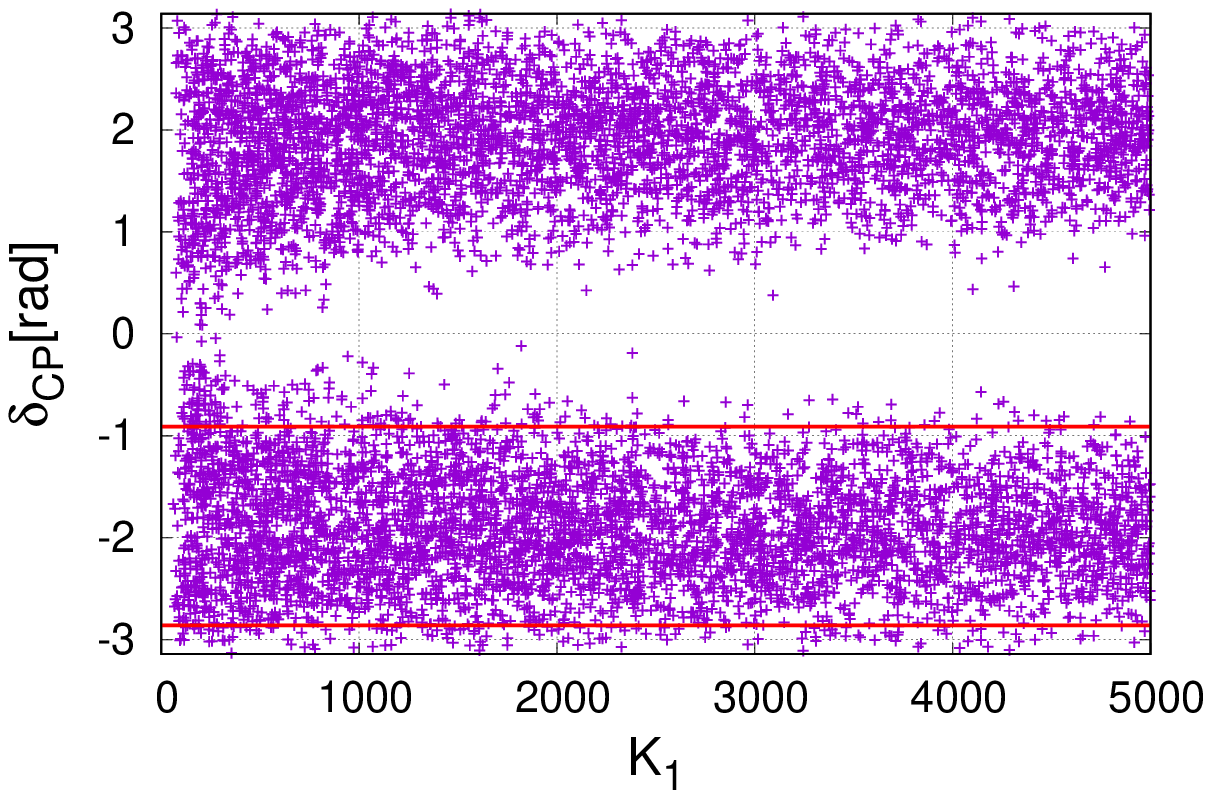}
\caption{The $K_1$ dependence of the predicted $\delta_{CP}$ at $K_1=1-5000$
by inputting the data of three mixing angles.
Here the horizontal red lines denote the NO$\nu$A experimental allowed region
 with $1\sigma$. }
\end{minipage}
\hspace{5mm}
\begin{minipage}[]{0.45\linewidth}
\includegraphics[width=8cm]{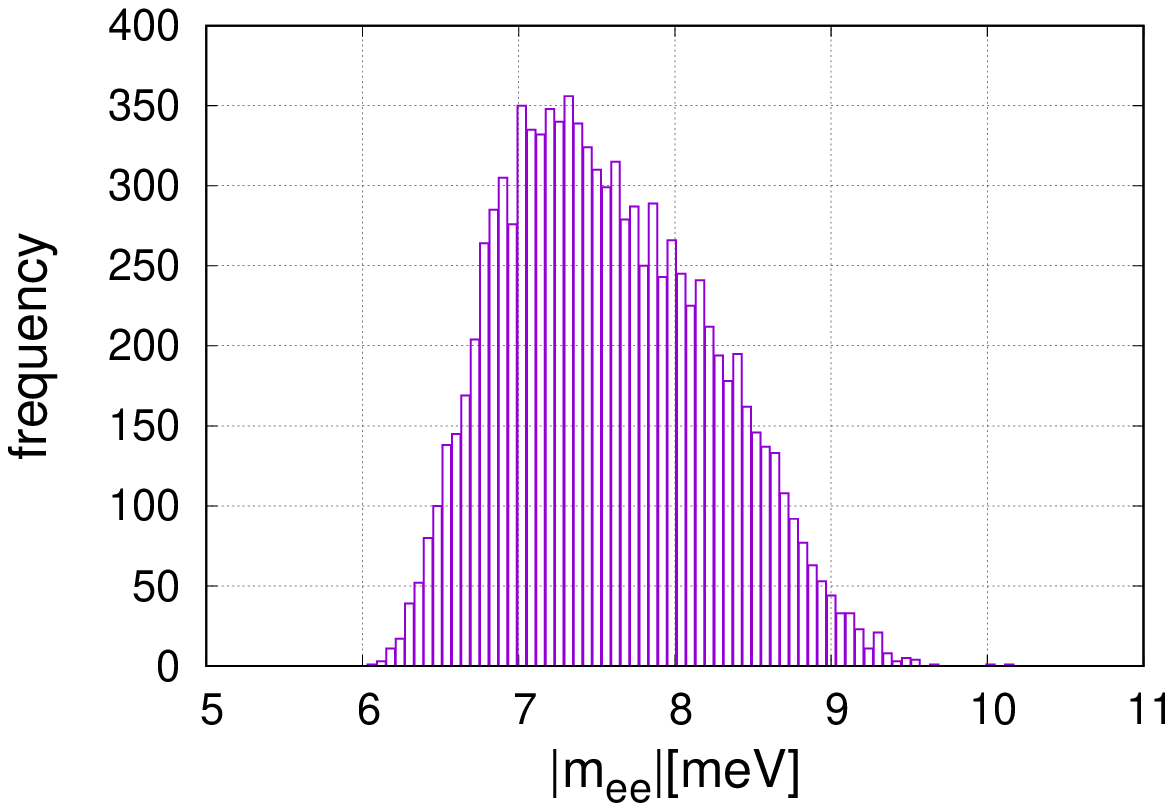}
\caption{The frequency distribution of the predicted $m_{ee}$ at $K_1=1-5000$ by inputting the data of three mixing angles.}
\end{minipage}
\end{figure}
\begin{figure}[b!]
\begin{minipage}[]{0.45\linewidth}
\includegraphics[width=8cm]{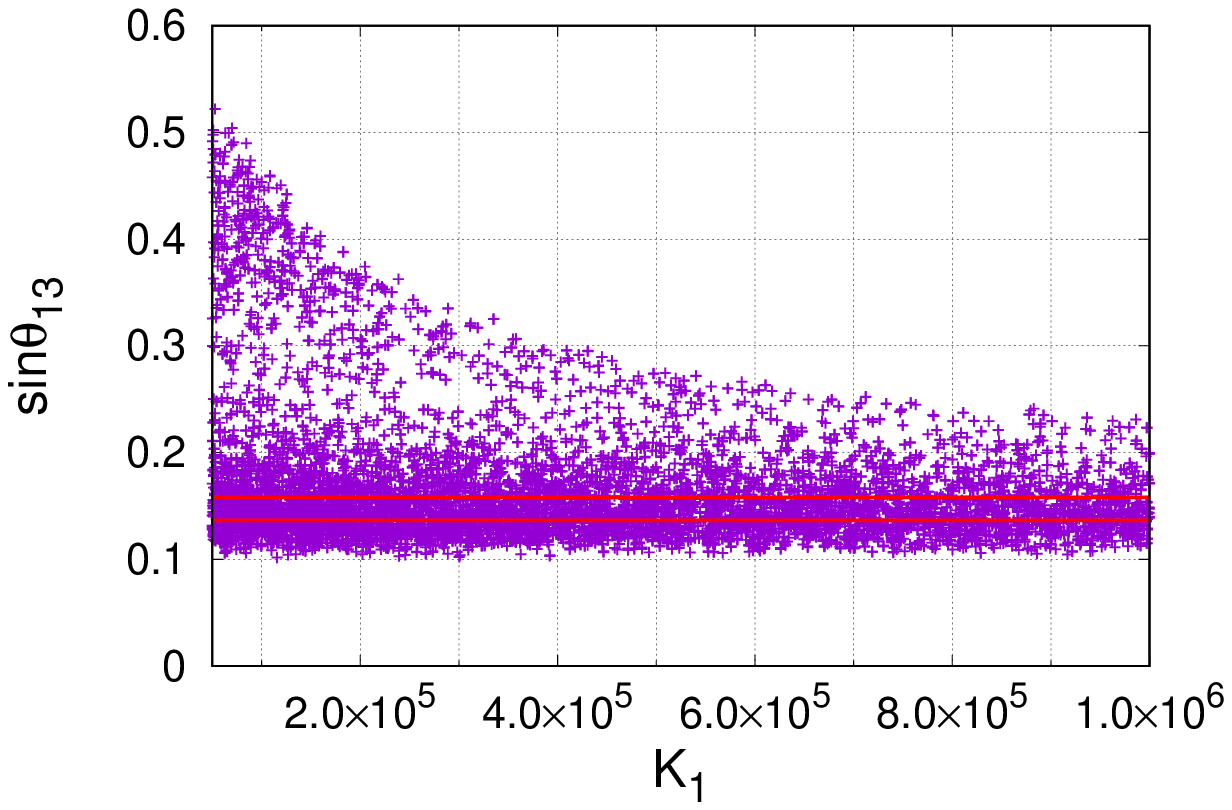}
 \caption{The $K_1$ dependence of the predicted $\sin\theta_{13}$ at $K_1=10^4 -10^6$
by inputting the data of  $\theta_{12}$ and $\theta_{23}$.
Here the horizontal red lines denote the experimental data with $3\sigma$. }
\end{minipage}
\hspace{5mm}
\begin{minipage}[]{0.45\linewidth}
\includegraphics[width=8cm]{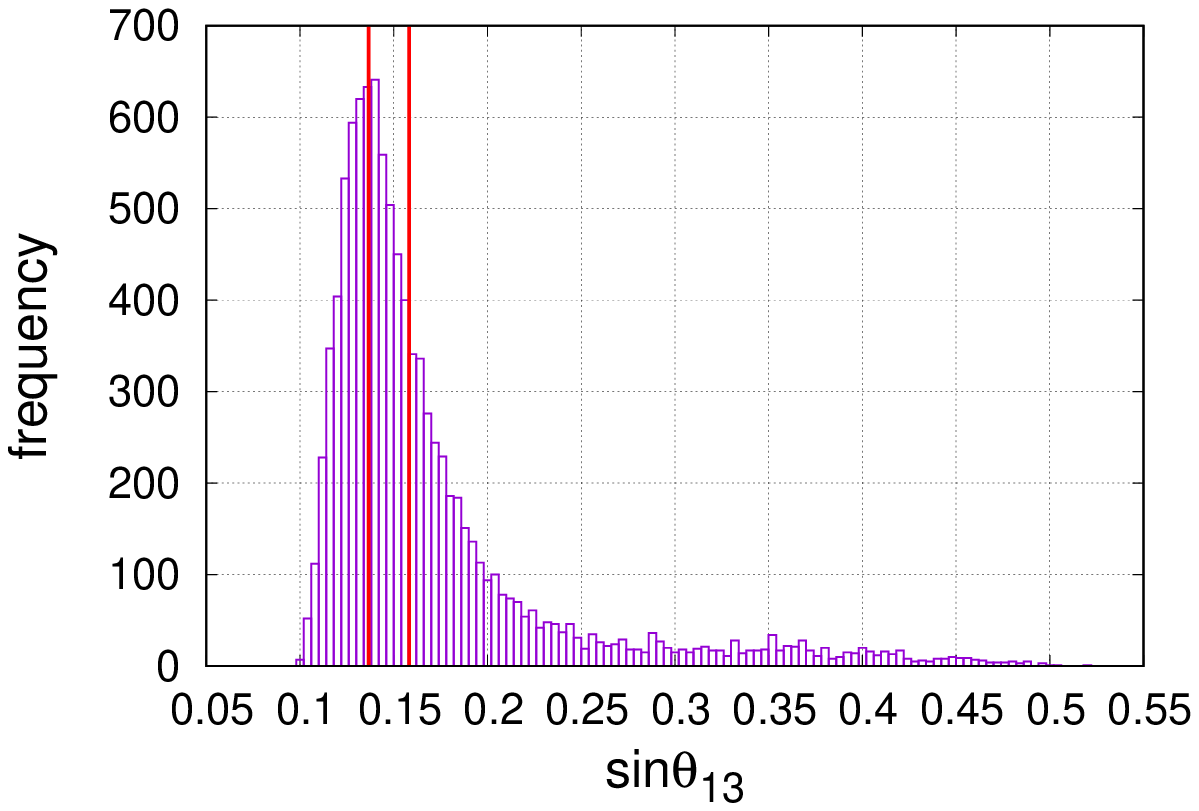}
\caption{The  frequency distribution of the predicted  $\sin\theta_{13}$ at $K_1=10^4 -10^6$
by inputting the data of  $\theta_{12}$ and $\theta_{23}$.
Here the vertical  red lines denote the experimental data  with $3\sigma$.}
\end{minipage}
\end{figure}

As shown in Fig.5, our result depends on the $K_1$.
Actually  the predicted region becomes narrow as $K_1$ increases significantly.
Let us discuss the result at  $K_1=10^4 -10^6$.
We show the  $K_1$ dependence of the predicted $\sin\theta_{13}$ at $K_1=10^4 -10^6$
by inputting the data of  $\theta_{12}$ and $\theta_{23}$ in Fig.7.
The mixing angle $\sin\theta_{13}$ is larger than $0.1$ in all region of $K_1$, but
the large mixing angle $0.5$  is allowed  below $K_1=10^5$.
However, it is remarked that  $\sin\theta_{13}$ decreases gradually and converges on the experimental allowed value.

In Fig.8,  we present  the frequency distribution of the predicted  $\sin\theta_{13}$ 
by inputting the data of the two mixing angles $\theta_{12}$ and $\theta_{23}$.
The distribution becomes rather sharp compared with the case of $K_1=1 -5000$.
The most favored region of  $\sin\theta_{13}$ is around $0.13-0.15$,
which is completely consistent with the experimental data.

\begin{figure}[t!]
\begin{minipage}[]{0.45\linewidth}
\includegraphics[width=8cm]{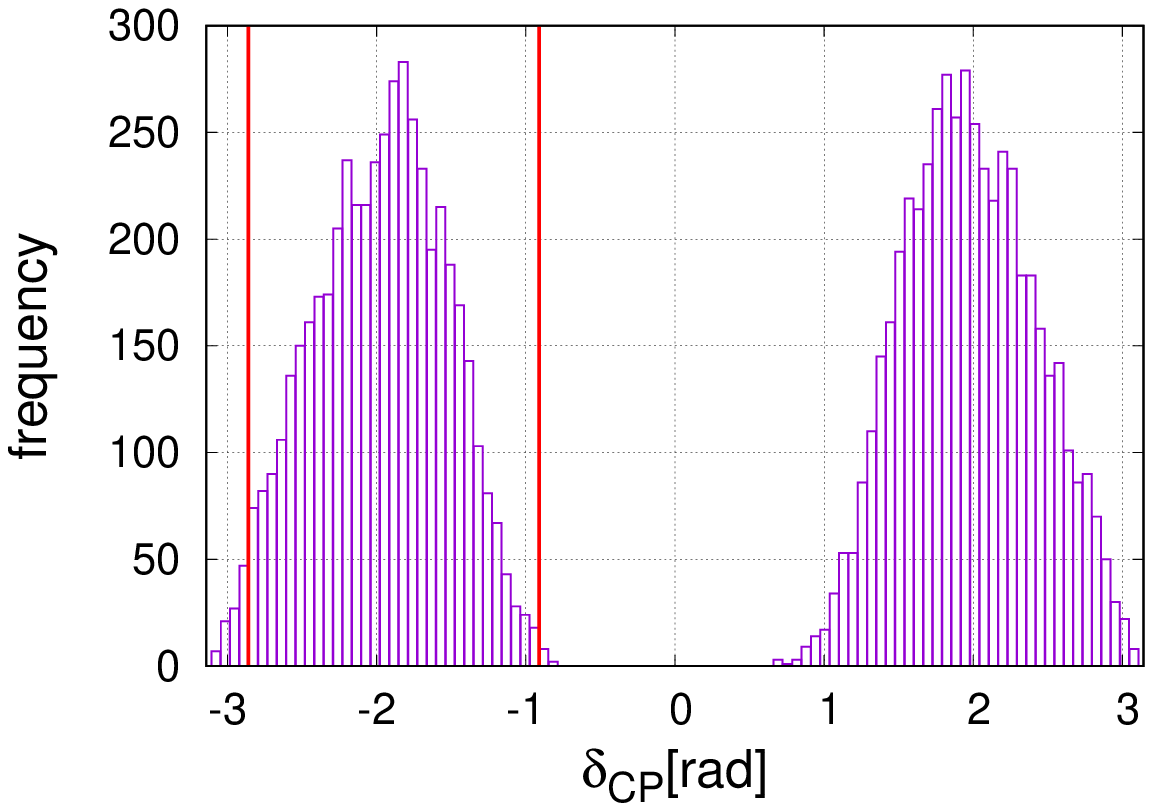}
\caption{The frequency distribution of the predicted $\delta_{CP}$ at $K_1=10^4 -10^6$
by inputting the data of three mixing angles.
Here  the vertical red lines denote the  NO$\nu$A  allowed region
 with $1\sigma$.}
\end{minipage}
\hspace{5mm}
\begin{minipage}[]{0.45\linewidth}
\includegraphics[width=8cm]{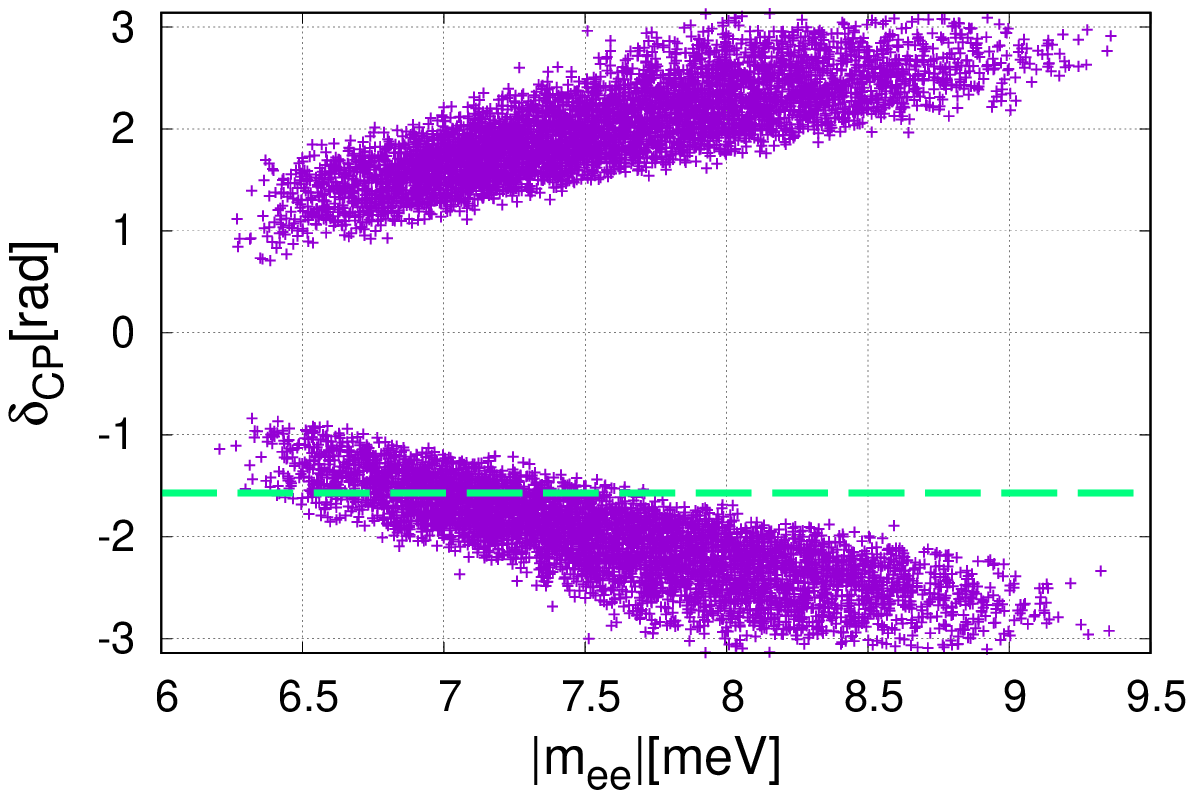}
\caption{The predicted  Dirac phase $\delta_{CP}$ versus the predicted $m_{ee}$
at $K_1=10^4 -10^6$ by inputting the data of three mixing angles.
Here  the horizontal green dashed line denotes  $\delta_{CP}=-\pi/2$ for the eye guide.}
\end{minipage}
\end{figure}
\begin{figure}[b!]
\begin{minipage}[]{0.45\linewidth}
\includegraphics[width=8cm]{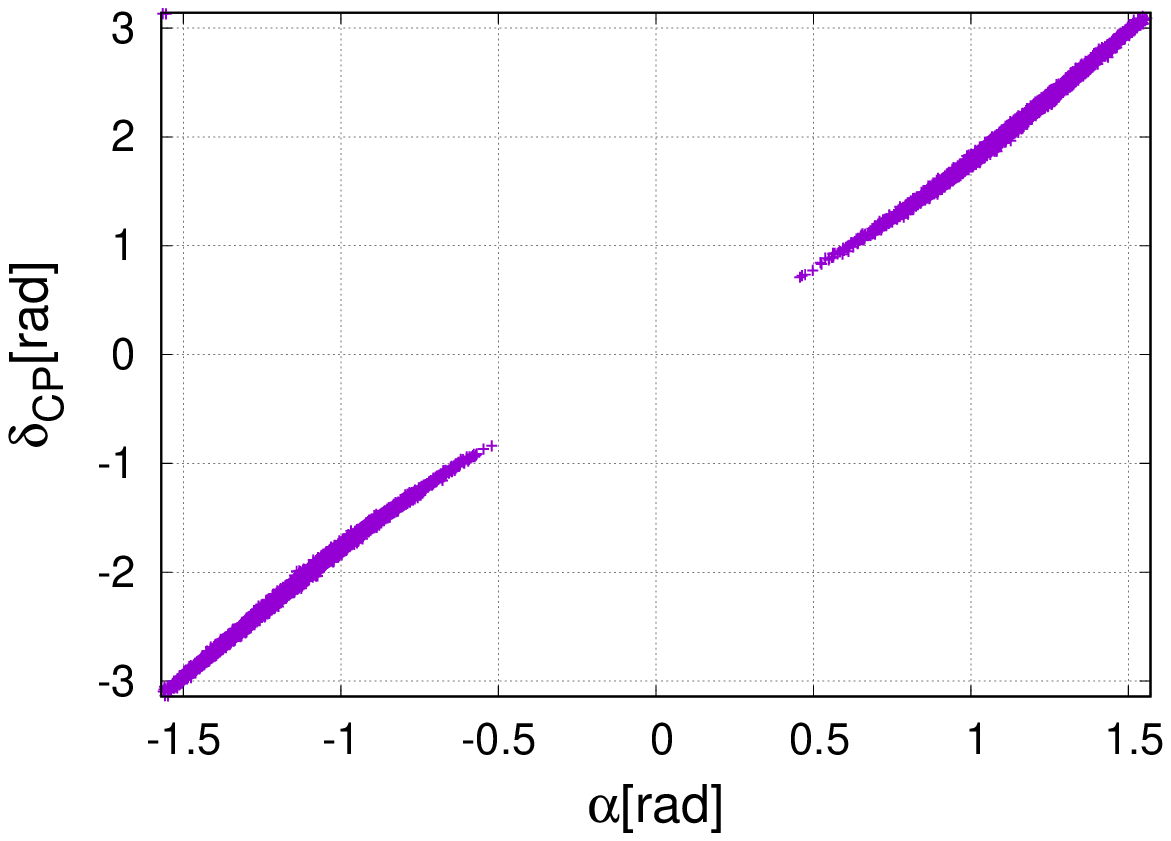}
\caption{The predicted  Dirac phase $\delta_{CP}$ versus the predicted Majorana phase $\alpha$
at $K_1=10^4 -10^6$ by inputting the data of three mixing angles.}
\end{minipage}
\hspace{5mm}
\begin{minipage}[]{0.45\linewidth}
\includegraphics[width=8cm]{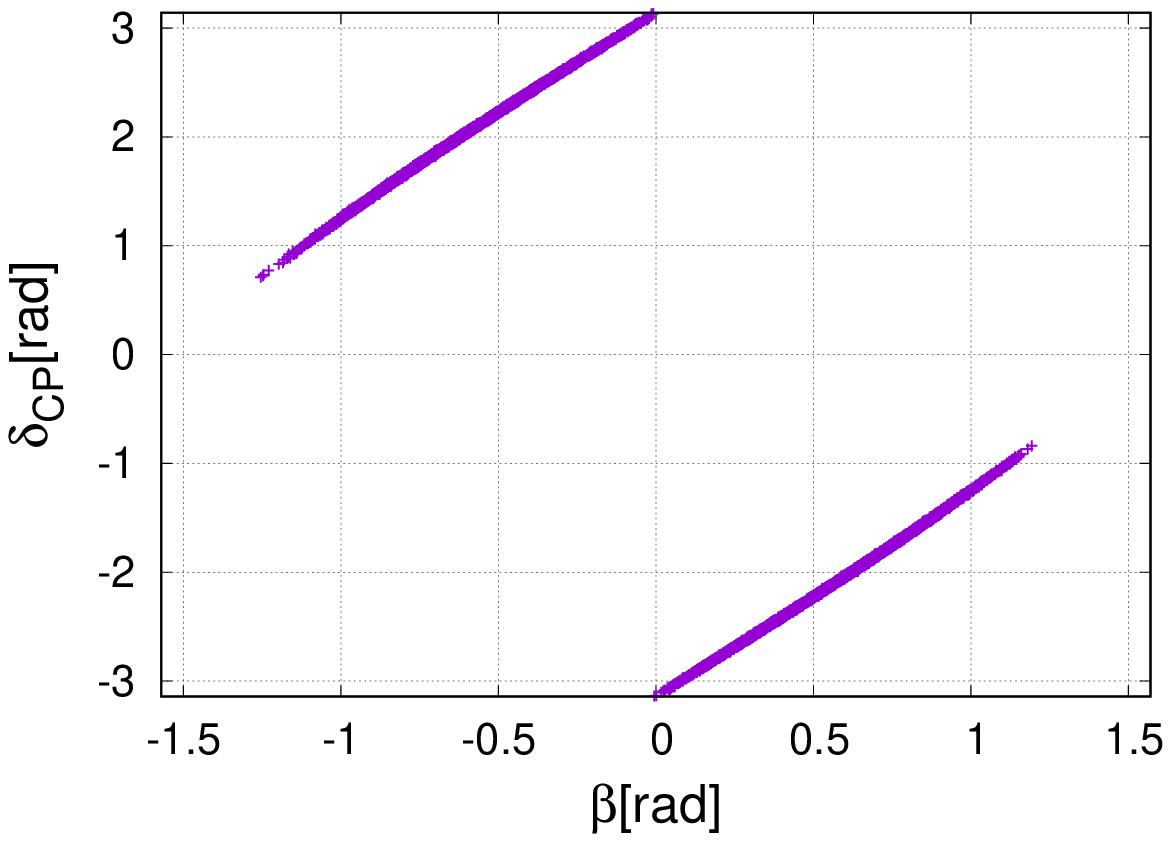}
\caption{The predicted  Dirac phase $\delta_{CP}$ versus the predicted Majorana phase $\beta$
at $K_1=10^4 -10^6$ by inputting the data of three mixing angles. }
\end{minipage}
\end{figure}

In Fig. 9,  we show the frequency distribution of the predicted value of $\delta_{CP}$
by inputting the data of the three mixing angles.
It is remarked that the peak of the frequency distributions of $\delta_{CP}$
 becomes close to $\pm \pi/2$.
Moreover, the region of  $\delta_{CP}=-1\sim 1$ radian is almost excluded.
Our result is consistent with the data of the T2K \cite{Abe:2013hdq} and 
the  NO$\nu$A \cite{Adamson:2016tbq} experiments.

The predicted  $m_{ee}$ of the neutrinoless double-beta decay is not so changed compared  
with the case of $K_1=1 -5000$. The favored  value of $m_{ee}$ is around $7\sim 8$ meV.
Here, we show the  predicted $\delta_{CP}$ versus  $m_{ee}$ by inputting the data of three mixing angles in Fig. 10. 
They are rather correlated as seen in Eq.(\ref{one-zero}) of Appendix.
If  $\delta_{CP}$ is restricted around $-\pi/2$ in the neutrino
experiment, the allowed region is restricted. Then, the predicted  $m_{ee}$ is  $6.5\sim 8$ meV.

\begin{figure}[t!]
\begin{minipage}[]{0.45\linewidth}
\includegraphics[width=8cm]{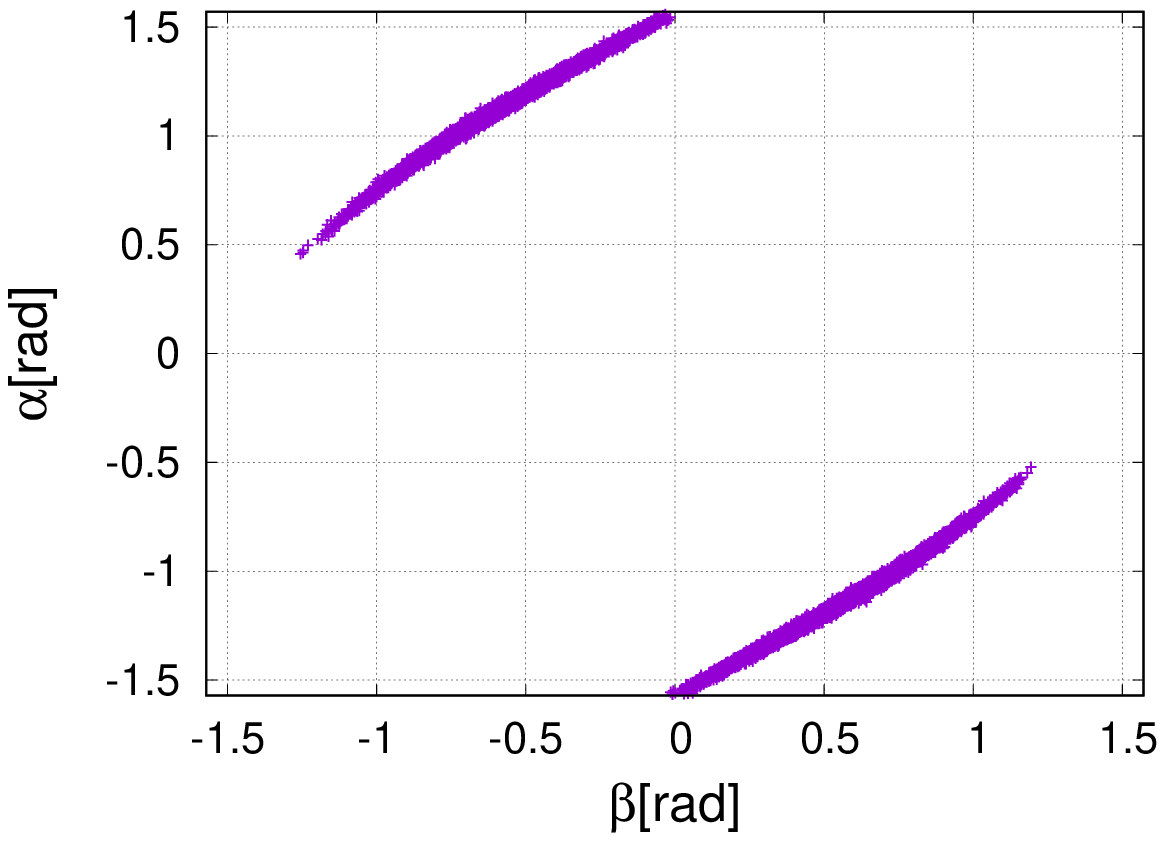}
\caption{The predicted  Majorana phase $\alpha$ versus the predicted Majorana phase $\beta$
at $K_1=10^4 -10^6$ by inputting the data of three mixing angles.}
\end{minipage}
\hspace{5mm}
\begin{minipage}[]{0.45\linewidth}
\includegraphics[width=8cm]{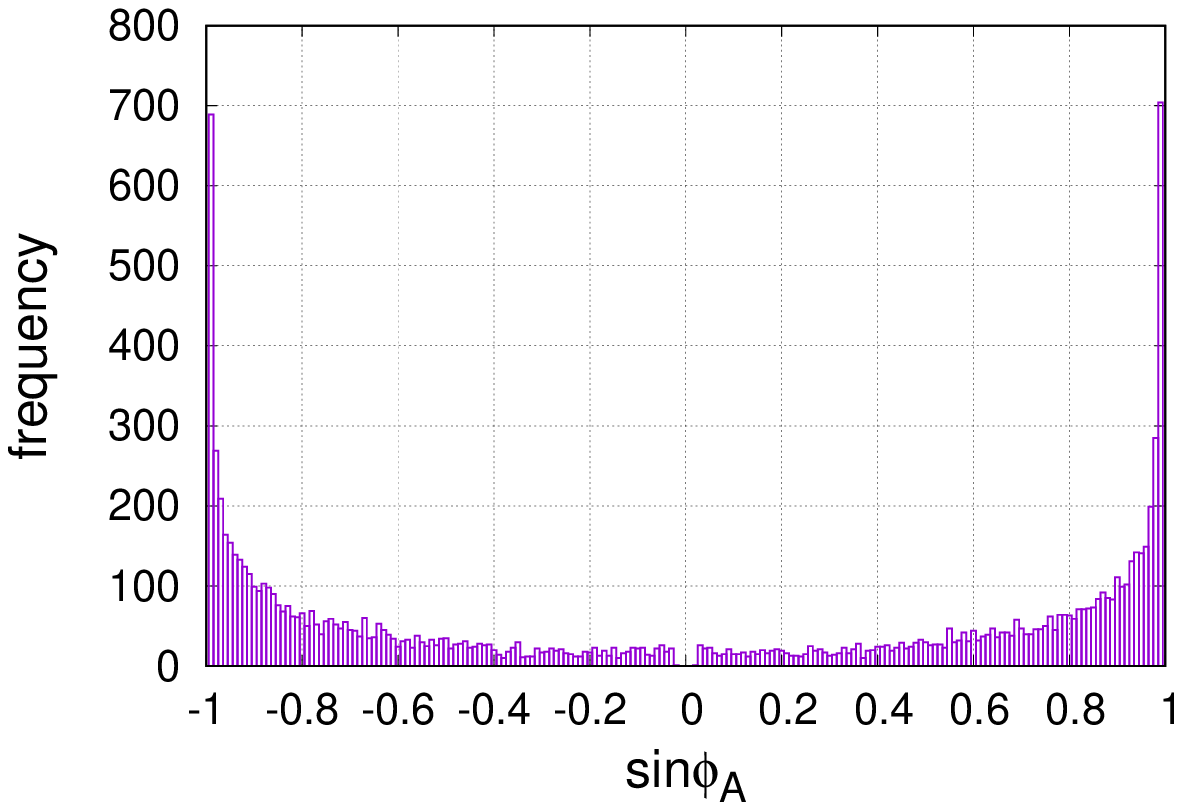}
\caption{The frequency distribution of the predicted $\sin \phi_A$ at $K_1=10^4 -10^6$
by inputting the data of three mixing angles. }
\end{minipage}
\end{figure}

At last, we show the correlation among the Dirac phase $\delta_{CP}$
and  the Majorana phases $\alpha$, $\beta$ in Figs. 11, 12 and 13.
There appears the tight correlation among them because we have only two phase parameters
in the neutrino mass matrix of Eq.(\ref{numatrix}).

\section{Summary}

We have presented the neutrino mass matrix  based on the Occam's Razor approach 
\cite{Harigaya:2012bw,Tanimoto:2016rqy}.  In the framework  of the seesaw mechanism,
we impose four  zeros in the Dirac neutrino  mass matrix, which give
the minimum number of parameters needed  for the observed neutrino masses and lepton mixing angles without assuming  any flavor symmetry.
Here, the charged lepton  mass matrix and the right-handed Majorana neutrino mass matrix
are taken to be  real diagonal ones.
Therefore,  the neutrino mass matrix  is given with  seven  parameters after absorbing
 the three phases into the left-handed neutrino fields.

Then, we obtain  the successful predictions of the mixing angle $\theta_{13}$  and
the CP violating phase $\delta_{CP}$ with the normal mass hierarchy of neutrinos.
We also discuss the Majorana phases $\alpha$ and  $\beta$ as well as the effective neutrino mass of the neutrinoless double-beta decay $m_{ee}$.
Especially, as $K_1$ increases to  $10^4\sim 10^6$, the predictions become more sharp.
The most favored region of  $\sin\theta_{13}$ is around $0.13\sim 0.15$,
which is completely consistent with the experimental data.
The  $\delta_{CP}$ is favored to be  close to $\pm \pi/2$, and the effective mass  $m_{ee}$ is around $7\sim 8$ meV.
 The reduction of the experimental error-bar  of the two mixing angles of $\theta_{12}$
 and $\theta_{23}$ will provide more precise predictions in our mass matrix of neutrinos.

Finally, it is emphasized that  we can perform a ``complete experiment"
to determine the low-energy neutrino mass matrix, since we have only
seven physical parameters in the mass matrix (see Eq.(\ref{numatrix})). In particular,
two CP violating phases $\phi_A$ and $\phi_B$ in the neutrino mass matrix are directly related to two CP violating phases at high energy. Thus, assuming the leptogenesis  we can
determine the sign of the cosmic baryon in the universe from  the low-energy experiments for the neutrino mass matrix. 
\footnote{The effect of quantum corrections of  the lepton mixing   matrix 
is neglected  in the evolution from the GUT scale to the electroweak scale 
for the normal mass hierarchy \cite{Haba:1999fk}.}
In fact the sign of baryon is given by the sign of $\sin \phi_A$
for the normal mass hierarchy $M_1< M_2 < M_3$ which is suggested from the predicted hierarchy $K_1>K_2>1$ shown in Fig.3.
 Unfortunately, the present experimental data show both sign allowed as shown in Fig.14.
\footnote{The detailed discussion on this issue will be given in the coming paper.}
We expect precise measurements of three mixing angles and  CP violating 
phases at low energy experiments.

\vspace{0.5 cm}
\noindent
{\bf Acknowledgement}

TTY thanks Prof. Serguey Petcov for the discussion on CP violation.
This work is supported by JSPS Grants-in-Aid for Scientific Research (No.28.5332; YS), Scientific Research (No.15K05045,16H00862; M.T) and
Scientific Research  (No.26287039,26104009,16H02176; TTY).
This work is also supported by World Premier International Research Center Initiative 
(WPI Initiative), MEXT, Japan.
YS is  supported in part by National Research Foundation of Korea (NRF) Research Grant NRF-2015R1A2A1A05001869.

\appendix
\section*{Appendix} 
By solving the eigenvalue equation in   Eq.(\ref{numatrix}),
the mass eigenvalues are expressed by  $a, b,c, K_1,K_2$ and $\phi_A, \phi_B$.
We have three equations among them as follows:
\begin{align}
  m_1^2+m_2^2&+m_3^2 \nonumber\\
  &= c^4 + b^4 (1 + K_2^2) + a^4 (K_1^2 + K_2^2) + 2 b^2 [a^2 (K_2^2 + K_1 \cos\phi_A) + c^2 (1 + K_2 \cos\phi_B)]
  \ , \nonumber \\
  \nonumber \\
  m_1^2 m_2^2+ &m_2^2 m_3^2+m_3^2 m_1^2 \nonumber\\
  &= b^8 K_2^2+a^8 K_1^2 K_2^2+2 a^6 b^2 K_1 K_2^2 (K_1+\cos\phi_A) \nonumber\\
  &+ a^4 [c^4 (K_1^2+K_2^2)+b^4 K_2^2 (1+K_1^2+4 K_1 \cos\phi_A) + 2 b^2 c^2 K_2 (K_2+K_1^2 \cos\phi_B)]\nonumber\\
  &+ 2 a^2 b^4 K_2 [b^2 (K_2+K_1 K_2 \cos\phi_A) + c^2 (K_2+K_1 \cos\phi_A\cos\phi_B + K_1 \sin\phi_A \sin\phi_B)]
  \ , \nonumber \\
  \nonumber \\
  m_1^2 m_2^2 m_3^2 &= a^8 c^4 K_1^2 K_2^2 \ .
  \label{massrelations}
\end{align}

Since the neutrino mass matrix in Eq.(\ref{numatrix}) has one zero, it constrains
the observed values.  Among three mixing angles, the three phases and the neutrino masses,
there is one relation:
\begin{align}
  0&=
  c_{12}c_{13}(-s_{12}c_{23}-c_{12}c_{23}s_{13}e^{i\delta_{CP}})e^{2i\alpha}m_1
  \nonumber\\
  &+
  s_{12}c_{13}(c_{12}c_{23}-s_{12}s_{23}s_{13}e^{i\delta_{CP}})e^{2i\beta}m_2
 +
  s_{13}s_{23}c_{13}e^{-i\delta_{CP}}m_3 \ .
  \label{one-zero}
\end{align}


\end{document}